\documentclass[10pt,a4paper,twoside]{aa}

\usepackage{graphicx}
\usepackage{natbib,twoopt}
\usepackage[breaklinks=true]{hyperref} 
 \bibpunct{(}{)}{;}{a}{}{,}    
 \newcommandtwoopt{\citeads}[3][][]{\href{http://adsabs.harvard.edu/abs/#3}%
                                        {\citealp[#1][#2]{#3}}}
 \newcommandtwoopt{\citepads}[3][][]{\href{http://adsabs.harvard.edu/abs/#3}%
                                        {\citep[#1][#2]{#3}}}
 \newcommandtwoopt{\citetads}[3][][]{\href{http://adsabs.harvard.edu/abs/#3}%
                                        {\citet[#1][#2]{#3}}}
 \newcommandtwoopt{\citeyearads}[3][][]%
   {\href{http://adsabs.harvard.edu/abs/#3}{\citeyear[#1][#2]{#3}}}
\bibpunct{(}{)}{;}{a}{}{,}
\usepackage{latexsym}
\usepackage{subfigure}
\usepackage{amsmath}
\usepackage{amssymb}
\usepackage{amstext}
\usepackage{color}
\usepackage{psfrag}
\usepackage{txfonts}

\def\ks{km s$^{-1}$~}
\def\d{$^\circ$}
\def\m{$^\prime$}
\def\s{$^{\prime\prime}$~}
\def\hh{$^{\mathrm h}$}
\def\mm{$^{\mathrm m}$}
\def\ss{$^{\mathrm s}$}
\def\cm3{cm$^{-3}$}

\def\12{$^{12}$CO}

\begin{document}
\title{The most complete and detailed X-ray view of the SNR Puppis~A}

\author{G. Dubner\inst{1}\fnmsep
\and 
N. Loiseau\inst{2,3}
\and P. Rodr\'{\i}guez-Pascual\inst{2}
\and M.~J.~S. Smith\inst{2,4}
\and E. Giacani\inst{1,5}
\and G. Castelletti\inst{1}
}

\offprints{G. Dubner}
\institute {Instituto de Astronom\'{\i}a y  F\'{\i}sica del Espacio (IAFE),UBA-CONICET,
CC 67, Suc. 28, 1428 Buenos Aires, Argentina\\
             \email{gdubner@iafe.uba.ar}
\and
\it XMM-Newton \rm Science Operations Centre, ESAC, Villafranca del Castillo, Spain
\and
ISDEFE, Madrid, Spain
\and
Telespazio Vega U.K. S.L.
\and
FADU, University of Buenos Aires, Buenos Aires, Argentina
}
 
   \date{Received <date>; Accepted <date>}


\abstract 
{}
 {With the purpose of producing the first detailed full view of Puppis  A in X-rays, we carried out new  \em XMM-Newton \rm observations covering the missing regions 
in the southern half of the supernova remnant (SNR) and combined them with existing \em XMM-Newton \rm and 
\em Chandra \rm data.
 }
  {Two pointings toward the south and southwest of Puppis~A were observed with \em XMM-Newton\rm. We  combined these data with archival \em XMM-Newton \rm and \em Chandra \rm data and produced images in the 0.3-0.7, 0.7-1.0 and 1.0-8.0 energy bands. 
  }
  {We present the first sensitive complete X-ray image of Puppis~A. We investigated  its
morphology in detail, carried out a multiwavelength analysis and estimated the flux density and luminosity of the whole SNR.
The complex structure observed across the  remnant confirms that Puppis~A evolves in an inhomogeneous, probably knotty interstellar medium. The southwestern corner includes filaments that perfectly correlate with radio features suggested to be associated with shock/cloud interaction. In the northern half of Puppis~A the comparison with \em Spitzer \rm infrared images  shows an excellent correspondence  between X-rays and 24 and 70~$\mu$m emission features, while to the south there are some matched and other unmatched features. X-ray flux densities of $12.6\times10^{-9}, 6.2\times10^{-9}$, and $2.8\times10^{-9}$~erg~cm$^{-2}$~s$^{-1}$ were derived for the 0.3-0.7, 0.7-1.0 and 1.0-8.0 keV bands, respectively. At the assumed distance of 2.2~kpc, the total X-ray luminosity between 0.3 and 8.0 keV is 1.2 $ \times 10^{37}$~erg~s$^{-1}$. We also collected and updated the broad-band data of Puppis~A between radio and GeV $\gamma$-ray range, 
producing its spectral energy distribution. To provide constraints to the high-energy emission
models, we re-analyzed radio data, estimating the energy content in 
accelerated particles to be  $\mathrm{U_{min}}=4.8 \times 10^{49}$~erg and the magnetic field strength 
$\mathrm{B}\sim 26$~$\mu$G.
 }
   {} 


\maketitle
\titlerunning{X-ray image of SNR Puppis~A}
\authorrunning{Dubner et al.}

\section{Introduction}

 Puppis~A is a nearby Galactic supernova remnant (SNR), about 50\m~ in diameter, 
 with age estimates ranging from 3700 yr \citep{winkler85} to 4450~yr \citep{becker12}.
It expands in the periphery of a large HI shell and near a chain of molecular
clouds that seem to surround most of the SNR \citep{dubner88}. Based on HI
and CO studies, a  distance of 2.2 kpc has been proposed for Puppis~A
\citep{reynoso03}, altough \citet{woermann00} suggested a closer distance of about
1.3 kpc based on 1667 MHz hydroxyl (OH) line detections.

Puppis~A is one of the brightest SNRs in X-rays. It has been observed
using the {\em Einstein} \citep{petre82}, {\em ROSAT} \citep{aschenbach93},
{\em Chandra} \citep{hwang05}, {\em Suzaku} \citep{hwang08}, and {\em  XMM-Newton \rm}
 \citep{hui06,katsuda10,katsuda12} satellites.
Based on the first images obtained with {\em Einstein}, \citet{petre82} concluded that on the large scale the X-ray surface
brightness increases from west to east, suggesting  a density gradient of a factor greater than 4 across the remnant. Subsequently, more sensitive observations showed a lower gradient in the 
emission distribution and confirmed that the SNR was highly structured, composed of a network of short, curved arcs.

Two bright knots were 
recognized early \citep{petre82}, one to the east, the ``bright eastern knot'' (BEK) and the other to the north. Spectral X-ray studies
provided evidence that the SNR shock has interacted with ambient clouds in a relatively late phase of evolution \citep{hwang05,
katsuda10, katsuda12} and complementary studies \citep{paron08,arendt10} indicated that the molecular cloud  has
already been destroyed by the shock in the BEK and that the X-ray and radio emission observed as bright features are, in fact, traces of
past shock/interstellar medium (ISM) interactions. Although most of the X-ray emission is dominated by the shocked ISM, there are also
isolated X-ray features rich in O, Ne, and Mg that can be identified as SN ejecta \citep{hwang08,katsuda08,katsuda10}.

In  optical wavelengths, the brightest filaments agree reasonably well on a large scale  with the location of radio and X-ray emission, but in detail
the appearance of the optical emission is notably different from the morphology displayed in X-rays, radio, and infrared.  A ridge of bright, nitrogen-rich
filaments (indicating that interstellar material is being shocked) is visible along the northwest shock front. Historically, this was recognized early by \citet{baade54} and \citet{goudis78}.
While these filaments move   
with velocities lower than $\sim$ 300 \ks, other set of O-rich filaments
with  chaotic morphology are detected with radial velocities higher than $\sim$ 1500 \ks. These last filaments are interpreted as knots of SN
ejecta that have remained relatively uncontaminated by interstellar material in spite of the thousands of years  since the explosion  \citep{winkler85}.

In the far-infrared domain, the first images of Puppis~A were obtained with the {\em IRAS} satellite. The SNR appeared bright and well resolved at 25 $\mu$m
and 60 $\mu$m \citep{arendt90}. New IR observations were carried out toward Puppis~A using the \em Spitzer \rm Space Telescope  MIPS photometer at 24, 70 and 160 $\mu$m \citep{arendt10}.
These new, high-quality data, revealed an extremely good IR/X-ray correlation on all spatial scales, confirming previous suggestions
 that the SNR's IR radiation is dominated by the thermal emission of swept-up interstellar dust collisionally heated by the hot  shocked gas.
 In particular, toward the BEK and the northern bright knot, the IR emission shows that the SNR shock has engulfed small denser clouds.

In radio, Puppis~A was observed using different facilities \citep{milne83, dubner91, castelletti06}. One remarkable
characteristic of Puppis~A in this spectral range is the lack of neat edges. The borders look quite structured, with many
short, arched filaments delimiting the remnant, particularly along the north-northeastern side  and to the southwest. These filaments
appear as narrow fringes whose spectrum alternates between flat and steep. This pattern has been  interpreted in
\citet{castelletti06} as a manifestation of Rayleigh-Taylor instabilities that distort the interface between ejecta and ambient
gas, stretching and compressing the magnetic field and thus modifying the observed synchrotron emission.

Close to the geometric center of Puppis~A lies the central compact object (CCO) RX~J0822-4300 that emits thermal X-rays with a harder spectrum than the surrounding medium. Similar
to other CCOs, it lacks  counterparts at any other wavelength as well as  a pulsar wind nebula (PWN). \citet{gotthelf09}
reported the discovery of pulsations with a periodicity of 112~ms and very peculiar emission properties. Several studies have
been conducted to determine the proper motion of RX J0822-4300 \citep{huibeck06, winkler07, deluca11,becker12}. Based on data
spanning a time baseline of more than ten years, \citet{becker12} concluded that the CCO has a proper motion of about 71~mas~yr$^{-1}$, which for a distance of 2~kpc implies a recoil velocity of about 670 \ks and an age of 4450 yr for the SNR.

The most recent discovery related to Puppis~A is the confirmed detection of Gev $\gamma-$ ray emission  with the {\em
Fermi} satellite \citep{hewitt12}. It is among the faintest SNRs yet detected at GeV energies. The high-energy source is
spatially extended. Its morphology was determined based on the {\em Fermi} LAT data which are restricted to events with energies above 800
MeV (to improve the angular resolution) and in the off-pulse window of the nearby (in the plane of the sky) Vela
pulsar. The GeV $\gamma-$ray emission extends over the northern half of the X-ray/radio remnant, with the maximum roughly
coincident with the location of the BEK. The $\gamma-$ray emission is well described by a power law with an index 2.1 in the
observed energy range. After  fitting the spectral energy distribution (SED) from radio to $\gamma-$rays using inverse Compton,
bremsstrahlung, and hadronic-dominated models,  \citet{hewitt12} concluded that any emission mechanism is possible with different magnetic field strengths and
different energetics of the relativistic particles.

Despite the rich nature of this remnant, whose emission  across the
 whole electromagnetic spectrum includes traces of the past star as well as
of the interaction of the SN blast wave with the surrounding gas, current X-ray images of
Puppis~A surprisingly lack a portion of the extended source.
The image constructed by \citet{petre82} from the combination of 11 {\em Einstein} HRI overlapping exposures was presented as an image of the entire extent of Puppis A. However, as noted by the authors, the coverage has two exceptions, the southeast and the southwest corners, which were not so far observed.
Moreover, the two southernmost regions observed 
happen to have the lowest exposure time of the entire source. Several years later, Puppis~A was observed using  {\em ROSAT} \citep{aschenbach93}. This image also failed to cover the southern region of Puppis~A, and almost 20\% of the surface of the SNR has not been observed in dedicated 
observations (note that the \em ROSAT \rm All Sky Survey includes an image of Puppis~A in its entirety, but with a poor angular resolution and sensitivity). The \em XMM-Newton \rm Slew Survey \citep{read2006}
also covered part of the southern regions of Puppis~A with irregular sensitivity, but good enough to
show that it was worth to make pointed observations of these regions. This 
incompleteness not only impedes  characterization of X-ray emission and prevent accurate estimates of the 
X-ray flux, but it makes  the interpretation of observations at other frequencies very difficult as well.
This motivated us  to carry out deep X-ray observations
towards the south and southwest portions of the remnant with the \em XMM-Newton \rm Observatory. We present here the first complete, detailed X-ray image of the Puppis~A SNR, including a sensitive study of the poorly
known southern region. A subsequent work will deal with the spectral
study based on the new observations.

\section{Observations and Data reduction}

\begin{figure}
\center
\vspace{1cm}
\includegraphics{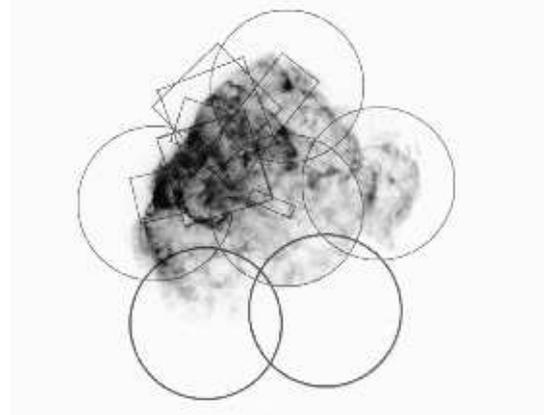}
     \caption{Largest previous detailed X-ray image obtained toward Puppis~A as obtained with 
\em ROSAT \rm HRI. Thin circles and squares represent the footprints of previous \em XMM-Newton \rm and {\em Chandra} observations, respectively. Thick circles show the footprints of the present observations.}
\label{footprints}
\end{figure}

To complete the X-ray view of Puppis~A, two new observations of the remnant
were performed with {\em XMM-Newton}. These pointings toward the south and southwest complement the existing coverage of the remnant obtained by the
{\em ROSAT}, {\em  XMM-Newton}, {\em Chandra}, and {\em Suzaku} observatories.
Figure~\ref{footprints} shows the footprints of the previous {\em XMM-Newton} and {\em Chandra} observations and, in addition, the two new fields presented in this paper.
The south and southwest fields were observed with the {\em XMM-Newton} EPIC pn and MOS cameras configured in full frame mode with thin filter. More details of these and the archival observations used in this paper are listed in Table 1.

{\em  XMM-Newton} data were reduced using the \em XMM-Newton \rm Science Analysis Software (SAS) version 12.0.1 with the most recent calibration files. High background flaring periods were filtered out using the SAS task espfilt, and images were subsequently accumulated using single and double events.

Archival {\em Chandra} observations of Puppis~A include data from both the ACIS and HRC instruments. However, observations from the latter are not included here because the instrument's intrinsic spectral resolution is too limited for the purposes of this paper. ACIS data were processed using CIAO version 4.4 and CALDB version 4.5.1, and images were accumulated following the standard CIAO data analysis threads\footnote{http://cxc.harvard.edu/ciao/}.

Images were created in four energy bands: 0.3 - 0.7 keV (covering OVII and OVIII 
emission), 0.7 - 1.0 keV (including Ne and FeL emission), 1.0 - 8.0 keV (including
 Mg, Si and S emission), and the wide band 0.3 - 8.0 keV.

The resulting 59 cleaned images (51 of {\em XMM-Newton} EPIC  and 8 of {\em Chandra} ACIS ) were 
each corrected for vignetting, weighted according to their respective energy-dependent effective areas, and subsequently merged with a 2-arcsec spatial binning. After smoothing with a 10-arcsec 
Gaussian kernel they were combined into the three-color image of the complete remnant, as shown in Fig.~\ref{rgb-xray}.

The previously best combination of Puppis~A images was done by \citet{katsuda10} using six
archival \em XMM-Newton \rm observations and five archival \em Chandra \rm observations.

\begin{table*}
\caption{\em XMM-Newton \rm and \em Chandra \rm observations}
\label{tab:param}
\centering
\begin{tabular}{l c c l l l l} 
\hline\hline
\multicolumn{7}{c}{\em XMM-Newton \rm  observations}\\ \hline
Field &  RA, Dec     &   Date   & \multicolumn{3}{l}{Net  exposure  time (sec)} &   observation ID  \\
& & & MOS1 & MOS2 & PN & \\
\hline
\multicolumn{7}{c}{{\bf New observations}}\\ \hline
 Puppis~A southwest &  08h21m20.00s -43d18'21.0" &  2012-05-05 &  35600 &  37403 & 26100 & 0690970101  \\
Puppis~A south    & 08h23m22.00s -43d21'00.0" &  2012-06-03 & 39900 &  39480 & 39360 &  0690970201 \\
\hline
\multicolumn{7}{c}{Archival observations}\\ \hline
PSR J0821-4300      &08h21m57.28s -43d00'17.5"&2010-05-02  &  1860 &  1020 &  6420 & 0650220201  \\
PSR J0821-4300      &08h21m57.28s -43d00'17.5"&2010-10-15  & 14600 & 13874 & 23040 & 0650220901  \\
PSR J0821-4300      &08h21m57.28s -43d00'17.5"&2010-10-15  & 22020 & 21286 & 16920 & 0650221001  \\
PSR J0821-4300      &08h21m57.28s -43d00'17.5"&2010-10-19  & 24640 & 25060 & 25210 & 0650221101  \\
PSR J0821-4300      &08h21m57.28s -43d00'17.5"&2010-10-25  & 20400 & 22820 & 21770 & 0650221201  \\
PSR J0821-4300      &08h21m57.28s -43d00'17.5"&2010-11-12  & 20380 & 22780 & 19570 & 0650221301  \\
PSR J0821-4300      &08h21m57.28s -43d00'17.5"&2010-12-20  & 24800 & 26660 & 13010 & 0650221401  \\
PSR J0821-4300      &08h21m57.28s -43d00'17.5"&2011-04-12  & 24600 & 28620 & 23970 & 0650221501  \\
RX J0822.0-4300     &08h21m57.40s -43d00'16.7"&2009-12-17  & 40550 & 36720 & 58300 & 0606280101  \\
RX J0822.0-4300     &08h21m57.40s -43d00'16.7"&2010-04-05  & 30060 & 26880 & 32460 & 0606280201  \\
Puppis~A west       &08h20m24.00s -42d55'13.0"&2005-10-09  &  8100 &  8100 &  5520 & 0303530101  \\
Puppis~A north knot &08h21m57.00s -42d36'58.0"&2003-04-17  &  7500 &  7800 &  6960 & 0150150101  \\
Puppis~A east knot  &08h24m11.00s -42d58'40.0"&2003-05-21  &  1440 &  1960 & 10900 & 0150150201  \\
Puppis~A east knot  &08h24m11.00s -42d58'40.0"&2003-06-25  &  5960 &  5640 &  4860 & 0150150301  \\
Puppis-A            &08h21m56.70s -43d00'19.0"&2001-11-08  & 10220 & 12400 & 20940 & 0113020301  \\
\hline
\hline
\multicolumn{7}{c}{\em Chandra \rm Observations}\\
\hline
Field   &  RA, Dec  &   Date   &  \multicolumn{3}{l}{Net  exposure  time (sec) } &   Observation ID  \\
& & & ACIS  &  \\
\hline
\multicolumn{7}{c}{Archival Observations}\\ \hline
Puppis~A NE filament      & 08h23m11.54s -42d52'09.12"  & 2010-11-13  &18000  &       & &13183   \\
Puppis~A NE filament      & 08h23m11.54s -42d52'09.12"  & 2010-11-10  &22000  &       & &12548   \\
Puppis~A NE shock front   & 08h23m08.16s -42d41'41.40"  & 2006-02-11  &28650 &       & &6371    \\
Puppis~A NE shock front   & 08h23m08.16s -42d41'41.40"  & 2005-09-04  &34480 &       & &5564    \\
PUPPIS N knot         & 08h22m00.00s -42d36'43.00" & 2002-03-09   &20150 &       & &1949 \\
PUPPIS bright E Knot      & 08h24m08.00s -42d57'00.00" & 2001-11-04   &10770 &       & &1951 \\
Puppis~A shock front      & 08h23m25.20s -42d44'21.30" & 2001-10-01   &15100  &       & &1950 \\
RX J0822-4300         & 08h21m57.50s -43d00'15.70" & 2000-01-01   &11940 &       & &750 \\
\hline
\end{tabular}\\
\end{table*}

\section{Results}

\begin{figure*}
\center
\vspace{1cm}
\includegraphics[width=\textwidth]{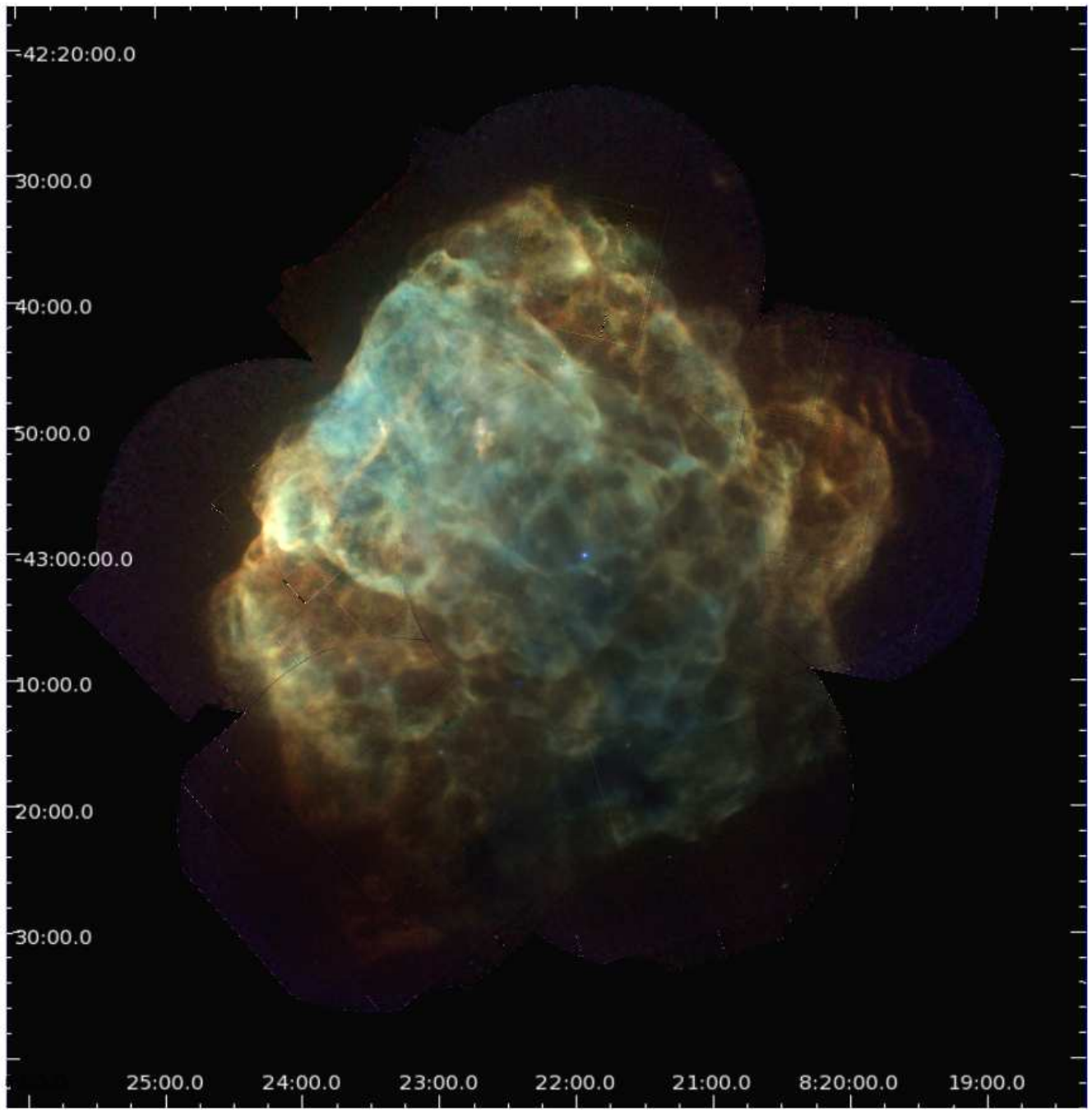}
\caption{Full view of Puppis~A in X-rays obtained after combining  two new {\sl XMM- Newton} observations with 23 existing  {\sl XMM- Newton} and {\em Chandra} pointings. In this representation, red, green, and blue correspond to  (0.3-0.7), (0.7-1.0) and (1.0-8.0) keV bands, respectively. The data are Gaussian-smoothed to an angular resolution of 10\s. The intensity scale is square root.  }
\label{rgb-xray}
\end{figure*}

Figure~\ref{rgb-xray}  displays a three-color image of Puppis~A.
This is the  most complete and detailed X-ray image of this extended SNR ever produced, including the southern regions that were not imaged in detail before. The appearance of Puppis~A in X-rays is quite unique. Although it contains the central compact object  RX J0822-4300  in its interior (the blue point), the X-ray emission is not dominated by a central pulsar wind nebula but is instead completely thermal in origin. The filamentary structure has a honeycomb appearance suggesting that the remnant is evolving in a rich and complex environment that strongly affects its morphology. From the comparison with high-quality X-ray images of other galactic SNRs, we find that this  ``cellular'' structure is similar to those observed  in the young SNRs Cas A, Crab, Kepler, and in G292.0+1.8, suggesting that we are seeing the imprint of the explosion on the ISM, even after about four thousand years. The new image of Puppis~A uncovered the existence of X-ray emission in the southwest corner with the same morphological pattern of curved filaments as seen in the rest of the SNR. 

\begin{figure}[h]
\centering
\includegraphics[width=0.5\textwidth]{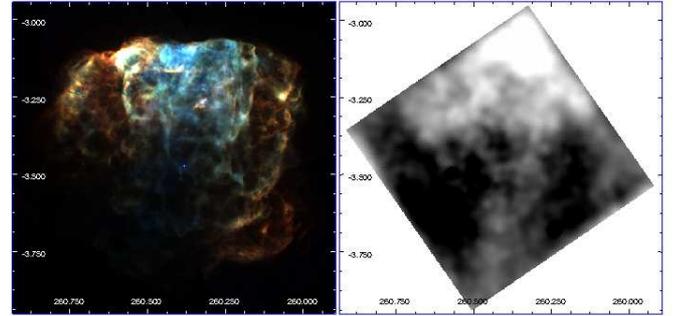}
   \caption{Comparison of the three-band X-ray emission (left) with the distribution of NH integrated between 0 and the systemic velocity of Puppis A (right), presented in Galactic coordinates. Note the central vertical band with higher column density in positional coincidence with the  blue fringe in the X-ray image.}
\label{nh}
 \end{figure}

From this  X-ray image of Puppis~A we can confirm that the bluish (harder X-ray emission) central band that crosses the remnant from 
northeast to southwest,  previously noticed by different authors \citep{aschenbach93,hwang05,katsuda10}, extends into the newly observed southwestern region. This peculiar harder band was first attributed to absorbing material related to the Vela SNR located between us and Puppis~A,  which would absorb most of the soft X-ray photons \citep{aschenbach93,hui06}. More recently,  \citet{katsuda10} proposed that it is more probably  caused by variations in the temperature and the ionization timescale than by variations in the intervening column density. To explore in more detail the question, we present in Fig. \ref{nh}  a comparison of the three-color X-ray image of Puppis A in Galactic coordinates (using a color display that emphasizes the harder spectrum blue band) with the HI column distribution as obtained from integrating the neutral hydrogen emission between 0 and 16.1 \ks, the systemic velocity of the SNR. The HI data were taken from the study conducted by \citet{reynoso03} in the direction to the central CCO, for which the authors combined interferometric HI observations performed using the Australian compact array ATCA, and single-dish observations carried out with Parkes 64-m telescope. There is a notable  higher-density column perpendicular to the Galactic plane direction, practically coincident in the plane of the sky with the blue strip,  suggesting that dense ISM might be responsible for absorbing soft X-ray photons. In addition, the 
wide-latitude CO survey of the third galactic quadrant carried out by \citet{may88} and the CO study by \citet{dame01} confirm at large scale the abundant  molecular gas below the Galactic plane (interpreted as a warp in the molecular disk), with small clouds along the line of sight to Puppis A  that appear to be roughly coincident with the HI observed band. A  detailed study of the atomic and molecular gas distribution across the whole SNR would be highly desirable to verify if differential absorption is responsible for the observed spectral variations.

\begin{figure*}[ht!]
\center
\vspace{1cm}
\includegraphics [width=\textwidth]{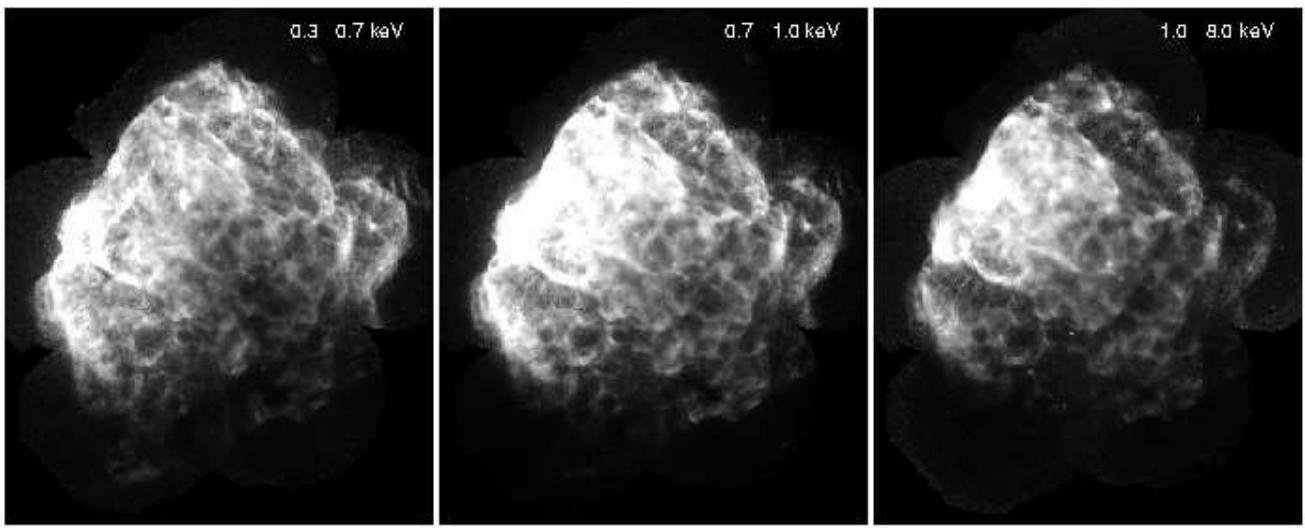}
     \caption{X-ray emission in the three observed bands. The grayscale is the same in the three images to allow a quick brightness comparison across the spectrum. Note in the hard X-ray  image (1.0 to 8.0 keV) that in addition to the central point source, the neutron star RX J0822-4300,  several point sources are now evident, particularly in the fainter southern half.}
\label{three-bands}
\end{figure*}

In Fig.~\ref{three-bands} we present the individual images in the soft, medium, and hard X-ray bands, where the grayscale was preserved to show the flux density changes across the X-ray spectrum. 
The SNR is clearly brightest in the band 0.7-1.0~KeV.

In addition to all the rich structure noticed  by other authors in the previous less 
sensitive X-ray images of the SNR,
the high dynamic range attained in this new image allows us to unveil the presence of  faint arcs and X-ray emitting extensions along most of the periphery,  features that have been overlooked before. Particularly  all along the northwestern border it can be noticed that the  ``scalloped''  view of the limb  exactly  corresponds with the  ``wave-like'' structures revealed in radio with  VLA observations at 1425 MHz \citep{castelletti06}. Moreover a series of faint, curved filaments with a concave shape can be seen ahead of the main shock to the west side of Puppis~A, near $(\alpha, \delta) = $ 08\hh 19\mm 30\ss, -42\d 50\m (ahead of the ``ear''). These features appear to be present only in the soft and medium energies. In contrast to these structured and/or blurred borders, some  portions of the outer rim are sharp, in some cases nearly linear while in other regions they have a concave geometry. Examples of these neat edges are observed along the northeast border around  
08\hh 23\mm 30\ss, -42\d 44\m,  ahead of  the position of  the BEK near 08\hh 24\mm 30\ss, -43\d 00\m, and in the newly mapped region, around 08\hh 21\mm 00\ss, -43\d 20\m, where the boundary  has a curious shape consisting of two arcs with a concave geometry. As we show in the  next section, these peculiar arcs coincide in location and shape with radio features. The southeast corner is probably the faintest part of the remnant. Some emission is detected in the 0.3 -0.7 keV band, while the medium-energy band is fainter and the hard X-ray emission is almost 
non-existent.
 
The high quality of the new hard X-ray image discloses several point sources in addition to the well-known central compact object RX J0822-4300. The three brightest sources visible in the southern half of Puppis~A are located at (08\hh 22\mm 27.0\ss, -43\d 10\m 
27.0$^{\prime\prime}$), (08\hh 22\mm 59.8\ss, -43\d 16\m 01.45$^{\prime\prime}$) and (08\hh 22\mm 32.8\ss, -43\d 21\m 01.9$^{\prime\prime}$). Of these, the first one has been listed in the Second \em XMM-Newton \rm Serendipitous Source Catalog as 2XMM J082226.9-431026 and in the Australia Telescope-PMN catalogue of Southern Radio Sources as ATPMN J082226.8-431026. Based on its radio spectrum, \citet{dubner91} confirmed its extragalactic nature. It is worthwhile to note that this is the only coincidence found between radio and X-ray point sources in the field of Puppis~A, though \citet{dubner91} reported the discovery of four compact sources across the surface of the SNR.

\subsection {X-ray and radio emission}

\begin{figure}[h]
\centering
\includegraphics[width=0.5\textwidth]{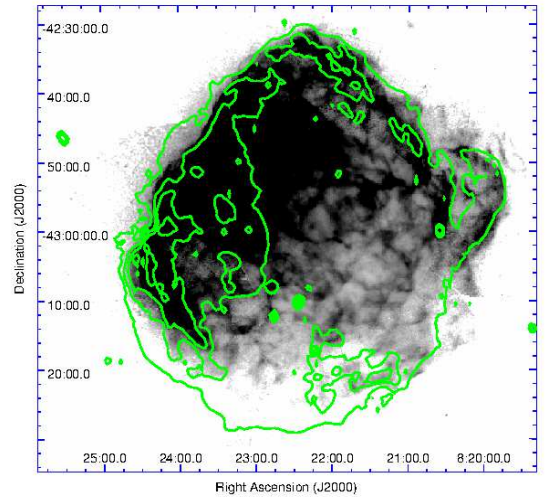}
\caption{X-ray emission between 0.7 and 1.0~keV in gray with overlapping radio contours 
(at 1425~MHz) as taken from \citet{castelletti06}. The plotted radio contours correspond to 4.5, 10, 20, and 50~mJy/beam, being the lowest contour six times above the rms noise.}
\label{xray-radio}
 \end{figure}

In Fig.~\ref{xray-radio} the X-ray emission in the medium energy (0.7-1.0~keV), the most intense band, is displayed (in gray) with a few radio contours  of the emission at 1425~MHz overlapped. \citet{castelletti06} analyzed  the comparison of radio with the X-ray emission in detail to the extent that was known then. Here we extend the previous analysis, including a careful study of the peripherical behavior in the northern half and of the weaker southern half. With this purpose in mind,  we saturated the grayscale to display the X-ray emission in  Fig.~\ref{xray-radio}.

All along the northeastern limb, it is remarkable how the sharp edge of the bright X-ray emission runs farther inside than the radio emission. This characteristic was noticed  by \citet{castelletti06} in the radio image of Puppis A, where a weak radio halo is apparent outside the bright radio rim. This diffuse radio emission might represent precursor synchrotron radiation from  relativistic electrons that have diffused upstream from the shock. The observed scale-length of the diffuse emission (approximately 2 arcmin $\simeq$ 1.3 pc at a distance of 2.2 kpc) agrees well with the typical upstream scale-length   for the  diffusive shock acceleration of cosmic rays by supernova remnants \citep{achterberg94}. On the other hand,  along  the north and west boundaries  radio and X-ray emissions  coincide in extension and in shape. This difference can be  explained by a change in the orientation of the magnetic field lines from almost perpendicular to the shock front on the northeastern side (where the scale-length of the halo attains its maximum) to almost parallel along the wave-like features on the northwestern side (where the scale-length is zero).

The appearance is different around the BEK  where the X-ray emission comes from evaporated clouds and, as discussed before by \citet{hwang05}, the observed indentation (both in X-rays and in radio) strongly suggests that the shock front has recently interacted with a dense obstacle and is wrapping it around.

  From  the analysis of the southern half of the SNR, two aspects can be noticed: toward the south and southeast,  the synchrotron radio emission extends considerably farther than the X-ray emission, while to the southwest the new image underscores the X-ray emission that exactly correlates with a bright radio feature. This radio maximum (near  08\hh 21\mm, -43\d 20\m) has  a flat radio spectral index \citep{castelletti06}, suggesting that this is  another site where the SN shock might be encountering dense interstellar material.
An inspection of the CO
data from the CfA CO survey \citep{dame01} reveals that in addition to the well-studied eastern cloud, there is another component at the same  kinematical distance  that appears to be almost in contact with
the southwestern features of enhanced radio emission (at least as seen with
the coarse angular resolution of the molecular data), suggesting that this may be another possible site
of shock/cloud interaction.

\subsection {X-ray and IR emission}

\begin{figure}[ht!]
\includegraphics[scale=3]{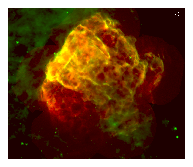}
\includegraphics[scale=3]{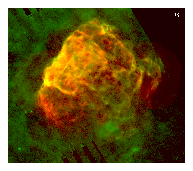}
\includegraphics[scale=3]{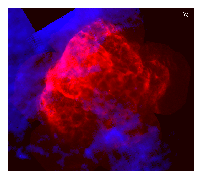}
\caption{
      Two-color image comparing  the X-ray emission between 0.7 and 1~keV (in red) with infrared emission as observed  with {\em Spitzer}  \citep{arendt10} at (\bf a\rm) 24~$\mu$m (in green), (\bf b\rm) 70~$\mu$m (in green), and (\bf c\rm) 160~$\mu$m (in blue). A color version of this figure is available in the online journal. } 
            \label{xray-ir-24}
            \end{figure}

Figure~\ref{xray-ir-24}\it a \rm  shows  a full view of  the correspondence between the X-ray emission in the 0.7-1.0 keV band  with the  infrared emission as detected with {\em Spitzer} at $\lambda$24~$\mu$m. This new comparison confirms the excellent agreement between X-rays and IR emission noticed by  \citet{arendt10} in the northern half of the remnant. To the southwest the new X-ray data allowed us to uncover a complex network of filaments that in general agree with the location of the 24 ~$\mu$m emission, although some X-ray features  lack an infrared counterpart and viceversa. While along  the northwestern flank an impressive match is observed between the small X-ray bright arcs  and dust emissions (seen in yellow in the online color version of the figure), to the southeast the newly detected X-ray emission  is not accompanied by analogous infrared emission.
The comparison with the IR  emission at 70~$\mu$m 
(Fig.~\ref{xray-ir-24}\it b\rm) shows characteristics similar to the former case.

The correlation with 160~$\mu$m emission 
(Fig.~\ref{xray-ir-24}\it c\rm) is very useful as a proxy to trace the spatial distribution of the foreground and co-spatial  interstellar gas. It corresponds to the larger size dust grains that are not destroyed by the passage of the shock front. The emission along the eastern flank seems to be associated with an external cloud that is being shocked by Puppis~A and partially covers its eastern border in the line of sight. This  cloud, rich in large dust grains, was not detected in the CO searches \citep{paron08} probably because the molecules have been  dissociated by the radiative precursor of the SNR. The dusty content of this dense material remained, however, and is clearly seen in the far-IR image.

 The IR-emitting region observed overlapping the south-southwest region corresponds exactly in shape and location with a minimum in the soft X-ray emission (see Fig.~\ref{rgb-xray}), demonstrating that it is a foreground cloud that absorbs the soft X-ray emission.  This IR feature  near 8\hh 22\mm.2, -43\d 21\m was previously reported by \citet{arendt10}, who identified it with a dark cloud located along the line of sight, closer than  1 kpc (corresponding to the  radial velocity of $\sim$ 6 \ks of associated molecular emission). Optical plates of the region reveal patchy obscuration to the south, confirming the presence of dense foreground gas.

\vspace{1cm}

\section {Physical properties}

Based on the new X-ray image, we determined the observed X-ray flux density
over the complete extent of Puppis A in soft, medium, and hard bands to within 5\%.
To determine the unabsorbed X-ray flux in these bands, we determined
correction factors to adjust the observed flux for the amount of interstellar
photoelectric absorption to the remnant. The correction is based on the spectral
fits of \citet{hwang08} to seven separate regions across Puppis A.
For each spectral fit, correction factors were
derived by determining the relative change in model flux in each of the three bands
when removing the interstellar absorption component from the model in question.
Of the set of factors thus obtained, the mean value was used as estimate of the global flux
correction factor for the respective energy band. From the range of factors we derived
the systematic error associated with the correction. The correction factors are
$8.3^{+7.9}_{-5.0}$ (soft), $2.6^{+0.7}_{-0.9}$ (medium) and $1.5^{+0.2}_{-0.2}$ (hard).
The resulting unabsorbed band flux-densities are presented 
in Table~\ref{tablesed} as a part
of the flux-density estimates across the whole electromagnetic spectrum (SED).  From these estimates,  the total X-ray flux measured between 0.3 and 8.0 keV is 
$ 21.6 ^{+14}_{-10}\times10^{-9}$~erg~cm$^{-2}$~s$^{-1}$, consistent within errors with previous estimates \citep [which were based on the {\em Einstein} HRI rate included in \citet{petre82} and referenced spectral parameters]{dwek87a}.

 At the assumed distance of 2.2~kpc the total luminosity in  X-rays is  $1.2\times10^{37}$~erg~s$^{-1}$. The X-ray luminosity can be compared with the radio luminosity  at 1425~MHz,  $\mathrm{L_{radio}}=1.5 \times 10^{34}$~erg~s$^{-1}$ \citep{castelletti06}, and in the FIR  $\mathrm{L_{IR}}=1.4 \times 10^{4}$ L$\sun$  $\sim 5 \times 10^{37}$~erg~s$^{-1}$ 
\citep{arendt10}. The luminosity ratio is
a useful parameter because it is independent of the distance. We find that  
$\mathrm{L_{X}}$ / $\mathrm{L_{radio}}$ = 800 and  $\mathrm{L_{X}}$ / $\mathrm{L_{IR}}$ = 0.24
(or $\mathrm{L_{IR}}$/$\mathrm{L_{X}}$ = 4.2).  This last ratio is similar to the value 5.2 estimated by \citet{dwek87b}.

\subsection{Spectral energy distribution}

For the sake of completeness, we present here a  compilation of the broad-band spectral energy distribution  with data homogenized to the same unities from radio to $\gamma$-rays, including for the first time data in the X-ray range as well as those reported in infrared by \citet{arendt10}.

\begin{figure}
\centering
\includegraphics[width=0.45\textwidth]{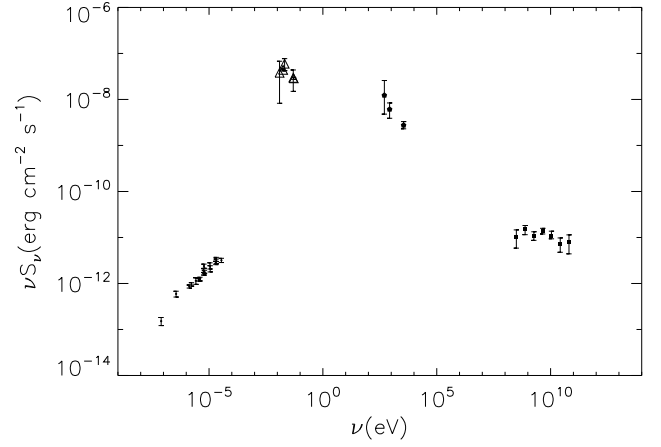}
\caption{Spectral energy distribution  of the SNR Puppis~A from radio to GeV energies.} 
\label{sed}
\end{figure}

Figure~\ref{sed} and Table~\ref{tablesed} show the SED for Puppis~A, where it is evident that the IR emission is the dominant radiative energy loss of the SNR. A similar behavior is  observed in the young SNR Cas A, where the infrared thermal emission from hot plasma emerges  as the main component in the global SED, lying much above the extrapolated radio synchrotron spectrum \citep{araya10}. 

\begin{table}[ht!]
\caption[]{Spectral energy distribution}              
\label{tablesed} 
\centering 
\begin{tabular}{lll}       
\hline\hline                        
\multicolumn{3}{c}{\bf Radio$^{\it a}$} \\ \hline
{\bf Frequency}& \bf Photon energy \rm & \bf Energy density $\nu S_{\nu}$ \rm \\
{\bf (MHz)} &     \bf (eV) \rm       & \bf  (erg~cm$^{-2}$~s$^{-1}$) \rm \\ 
\hline
19 &   7.92 $\times$ 10$^{-8}$  & (15.20 $\pm$ 3.04) $\times$10$^{-14}$ \\
86 &   3.58 $\times$ 10$^{-7}$  & (59.34 $\pm$ 8.60) $\times$10$^{-14}$ \\
327 &  1.36 $\times$ 10$^{-6}$  & (86.00 $\pm$ 6.54) $\times$10$^{-14}$ \\
408 &  1.70 $\times$ 10$^{-6}$  & (95.88 $\pm$ 8.16) $\times$10$^{-14}$ \\
635 &  2.64 $\times$ 10$^{-6}$  & (11.43 $\pm$ 1.84) $\times$10$^{-13}$ \\
843 &  3.51 $\times$ 10$^{-6}$  & (12.14 $\pm$ 0.84) $\times$10$^{-13}$ \\
960 &  4.00 $\times$ 10$^{-6}$  & (12.48 $\pm$ 1.15) $\times$10$^{-13}$  \\
1410 & 5.87 $\times$ 10$^{-6}$  & (18.19 $\pm$ 2.82) $\times$10$^{-13}$ \\
1425 & 5.94 $\times$ 10$^{-6}$  & (16.25 $\pm$ 1.14) $\times$10$^{-13}$ \\
1440 & 6.00 $\times$ 10$^{-6}$  & (23.90 $\pm$ 2.45) $\times$10$^{-13}$ \\
1515 & 6.31 $\times$ 10$^{-6}$  & (17.88 $\pm$ 1.52) $\times$10$^{-13}$ \\
2650 & 1.10 $\times$ 10$^{-5}$  & (24.38 $\pm$ 3.71) $\times$10$^{-13}$ \\
2700 & 1.12 $\times$ 10$^{-5}$  & (21.06 $\pm$ 3.24) $\times$10$^{-13}$ \\
4750 & 1.98 $\times$ 10$^{-5}$  & (28.03 $\pm$ 2.38) $\times$10$^{-13}$ \\
5000 & 2.08 $\times$ 10$^{-5}$  & (33.50 $\pm$ 3.50) $\times$10$^{-13}$ \\
5000& 2.08 $\times$ 10$^{-5}$  & (30.50 $\pm$ 3.50) $\times$10$^{-13}$ \\
8400 & 3.50 $\times$ 10$^{-5}$  & (31.92 $\pm$ 3.36) $\times$10$^{-13}$ \\
\hline
\multicolumn{3}{c}{\bf Infrared\tablefootmark{b}} \\ \hline
{\bf Wavelength}  & \bf Photon energy \rm & \bf Energy density $\nu S_{\nu}$ \rm \\
{\bf ($\mu$m)} &     \bf (eV) \rm       & \bf  (erg~cm$^{-2}$~s$^{-1}$) \rm \\  
\hline
100 & 1.25 $\times$ 10$^{-2}$ & (3.80 $\pm$ 2.97)  $\times$10$^{-8}$     \\
70 & 1.79 $\times$ 10$^{-2}$ &  (4.40 $\pm$  0.22) $\times$10$^{-8}$        \\
60 & 2.08 $\times$ 10$^{-2}$ &   (5.98 $\pm$  1.75) $\times$10$^{-8}$        \\  
25& 4.99 $\times$ 10$^{-2}$  &   (2.94 $\pm$ 1.44) $\times$10$^{-8}$          \\
24& 5.20 $\times$ 10$^{-2}$  &    (2.88 $\pm$ 0.14) $\times$10$^{-8}$  \\
\hline
\multicolumn{3}{c}{\bf X-rays\tablefootmark{c}} \\ \hline
{\bf Energy}  & \bf Photon energy \rm & \bf Energy density $\nu S_{\nu}$ \rm \\
{\bf (keV)} &     \bf (eV) \rm       & \bf  (erg~cm$^{-2}$~s$^{-1}$) \rm \\  
\hline
 &&\\
 $[$0.3-0.7$]$ & 0.50 $\times$ 10$^{3}$ & $12.6^{+13.2}_{-7.8}\times10^{-9}$    \\
 &&\\
 $[$0.7-1.0$]$& 0.85 $\times$ 10$^{3}$ & $ 6.2^{+2.2}_{-2.3}\times10^{-9} $ \\ 
 &&\\
 $[$1.0-8.0$]$ & 3.50 $\times$ 10$^{3}$ & $ 2.8^{+0.5}_{-0.5}\times10^{-9} $   \\
  &&\\
\hline
\multicolumn{3}{c}{\bf $\gamma$-rays\tablefootmark{d}} \\ \hline
 & \bf Photon energy \rm & \bf Energy density $\nu S_{\nu}$ \rm \\
 &     \bf (eV) \rm       & \bf  (erg~cm$^{-2}$~s$^{-1}$) \rm \\  
\hline
 & 312 $\times$ 10$^{6}$ &  (10.0 $\pm$ 5.1) $\times$10$^{-12}$ \\
 & 757 $\times$ 10$^{6}$ &   (15.0 $\pm$ 4.1) $\times$10$^{-12}$ \\
 & 1841 $\times$ 10$^{6}$ &  (11.0 $\pm$ 2.4) $\times$10$^{-12}$  \\
 & 4472 $\times$ 10$^{6}$ & (14.0 $\pm$ 2.1) $\times$10$^{-12}$  \\
 & 1086.6 $\times$ 10$^{7}$ & (11.0 $\pm$ 2.4) $\times$10$^{-12}$  \\
 & 2640.3 $\times$ 10$^{7}$ & (7.3$ \pm$ 2.5) $\times$10$^{-12}$  \\
 & 6415.3 $\times$ 10$^{7}$ & (7.8 $\pm$ 2.6) $\times$10$^{-12}$  \\
\hline  
\end{tabular}
\tablefoot{
\tablefoottext{a}{from \citet{castelletti06}};
\tablefoottext{b}{from \citet{arendt10}};
\tablefoottext{c}{this work};
\tablefoottext{d}{from \citet{hewitt12}} }
\end{table}

\subsection{Energetics and magnetic field strength}

In their analysis of the {\em Fermi} LAT GeV observations of Puppis~A, \citet{hewitt12} concluded that by adopting different magnetic field strengths, all emission mechanisms are able to fit the data and both leptonic and hadronic-dominated models can reproduce the non-thermal SED that they presented, leaving the question about the origin of the GeV emission still open.  The authors estimated that a total content of cosmic-ray accelerated electrons and protons of  at least 1 to 5 $ \times 10^{49}$~erg is required to produce the observed $\gamma-$ray emission. Although the search for the mechanism responsible for the high-energy emission is beyond the scope of this paper, we can set observational constraints that may help to fit the models in the future. To accomplish this, we re-analyzed the VLA radio continuum data presented by \citet{castelletti06} to  estimate  the minimum energy  content of the accelerated particles and the corresponding B field.

From the synchrotron emissivity it is not possible to derive  the magnetic field value unambiguously. The usual way of estimating the magnetic field strength in a radio source is to minimize its total energy content. The total synchrotron energy observed is stored in the magnetic field as well as in the kinetic energy of the relativistic particles (electrons and baryons).  If we adopt the  hypothesis of equipartition between particles and magnetic energy, the minimum energy  content and the corresponding magnetic field can be obtained from the relations presented by \citet{mof75}:  
$\mathrm{U_{min}}=0.50\, (a\,A\,L)^{4/7}\,V^{3/7}$ and 
$\mathrm{B}=2.3\,(a\,A\,L/V)^{2/7}$, 
where $V$ is the volume of the source, $A$ is a factor
that depends on the assumed lower and upper cutoff radio frequencies and
the radio spectral index of the SNR, and $L$ is the total luminosity of the
source in radio wavelengths. As mentioned above,  electrons
may not be the only energetic particles within the source; an
appreciable amount of energy may also be stored in energetic baryons.
The relative amount can be described as $\mathrm{U_{p}}=a\,\mathrm{U_{e}}$,
where $a$  has been estimated by various authors to vary between 1 and 100 \citep{mof75}.
If we assume a=50,  for the typical lower and upper cutoff frequencies
of 10$^{7}$ and 10$^{11}$ MHz, we derive for Puppis~A a minimum energy content of about  
$\mathrm{U_{min}}=$ 4.8 $\times$ 10$^{49}$~erg. The magnetic field for which the total energy content is minimum is  $\mathrm{B_{min}} \sim$ 26 ~$\mu$G. These values agree well with the results derived by \citet{hewitt12} from gamma-ray model fiting, particularly with the estimates from the hadron-dominated model. It is also comparable with the energy content derived for Cas A \citep{araya10}.

\section {Summary}

Based on our new \em XMM-Newton \rm observations and the combination with archival data acquired with the same telescope and with {\em Chandra}, we have produced the highest quality, most comprehensive X-ray image ever made of the SNR Puppis~A  in the energy bands 0.3-0.7~keV, 0.7-1.0~keV and  1.0-8.0~keV. The high dynamic range attained in this new image revealed weak emission  not only in the newly explored southern region, but also all  along the periphery
of the SNR. The new image shows for the first time X-ray emission in the south/southwest parts of Puppis~A. In these areas the emission is in the form of short arched filaments, similar in appearance to the rest of the source.
The southwestern border has a  peculiar morphology consisting of two arcs with concave geometry, coincident with radio features with properties suggestive of shock/cloud interaction. The new image also uncovered several hard-spectrum point sources, one of which was previously identified in the radio band as extragalactic. The new observations also confirmed that a broad band with a harder X-ray spectrum, which crosses the SNR from northeast to southwest and was shown in previous studies of this remnant, extends into the southwest corner. An inspection of the HI distribution in the central part of Puppis~A shows a band with a higher column density in positional coincidence with the observed hard X-ray feature, suggesting that it might be  the consequence of localized absorption of soft photons by intervening gas.

The new image was compared with a VLA radio image at 1425 MHz. Of particular interest is the fact that to the east the radio limb is about 2 arcmin  ($\sim$ 1.3~pc at the distance of~2.2 kpc) outside the sharp X-ray border.  We interpreted this  radio halo as the precursor synchrotron radiation from  relativistic electrons that have diffused upstream from the shock. The scale-length is comparable with the theoretically estimated length for the diffusive shock acceleration of cosmic rays by supernova remnants.  
To the north and northwest,  radio and X-ray borders show a notable coincidence both in shape and location. To the south the agreement is poorer, except to the southwest, as mentioned above.

From the comparison with {\em Spitzer} infrared images at 24, 70 and 160~$\mu$m we were able to confirm the previously reported excellent IR/X correspondence in the northern half of the remnant. The correlation is poorer toward the south, where there are X-ray features without an IR counterpart, and viceversa. The comparison with the FIR image at 160~$\mu$m
showed a broad strip of IR emission all along the eastern flank of Puppis~A, abutting the X-ray emission. We believe that this emission comes from the bordering molecular cloud, where the large-scale dust grains have not been destroyed by the SN shock.

Moreover, based on the new image, we carefully estimated the flux density as observed in the three energy bands in 12.6, 6.2, and 2.8 $\times 10^{-9}$~erg~cm$^{-2}$~s$^{-1}$ for soft, medium, and hard X-ray band, respectively. At the assumed distance of 2.2~kpc, the total X-ray luminosity between 0.3 and 8.0 keV is 1.2 $ \times 10^{37}$~erg~s$^{-1}$, which is 800 times higher than the radio luminosity and only a quarter of the infrared luminosity, confirming the significance of the IR emission in the energetics of Puppis~A.

Additionally,   to set constraints that may help understanding the mechanisms responsible for the GeV emission, we presented an updated and complete broad-band SED of Puppis~A, including the X-ray fluxes from this work. We also re-analyzed radio data to estimate the minimum energy content stored in relativistic particles, as well as the magnetic field strength. From these calculations we derived  $\mathrm{U_{min}} = 4.8 \times 10^{49}$~erg and $\mathrm{B} \sim 26~\mu$G.

A forthcoming paper will deal with the  spectral study of Puppis~A, making use  of the new observations reported here.

\vspace{1cm}

\begin{acknowledgements}
We are very grateful to  R.~D. Saxton and A.~M. Read for their valuable contribution during the first stages of this work.
We acknowledge J. Hewitt for providing the {\em Fermi} GeV data and to R. Arendt for the {\em Spitzer} IR images and for helpful comments.
E.G. acknowledges support
from the Faculty of the European Space Astronomy Centre (ESAC). GD acknowledges the hospitality during her stay at ESAC.
This research was
partially funded by Argentina Grants awarded by
 ANPCYT: PICT 0902/07, 0795/08 and 0571/11; CONICET: 2166/08, 0736/12 and University of Buenos Aires (UBACYT 20020100100011).
This work is based on observations
done with the \em XMM-Newton \rm, an ESA science mission with instruments
and contributions directly funded by ESA Member States and the US (NASA). G.D, E.G. and G.C. are Members of the Carrera del Investigador Cient\'\i fico of CONICET, Argentina.

\end{acknowledgements}

\bibliographystyle{aa}  
\bibliography{bib-puppis}
\IfFileExists{\jobname.bbl}{}
{\typeout{}
\typeout{****************************************************}
\typeout{****************************************************}
\typeout{** Please run "bibtex \jobname" to optain}
\typeout{** the bibliography and then re-run LaTeX}
\typeout{** twice to fix the references!}
\typeout{****************************************************}
\typeout{****************************************************}
\typeout{}
}

\end{document}